\RequirePackage{lineno}
\documentclass[12pt]{iopart}
\usepackage{hyperref}
\usepackage{graphicx}
\usepackage{gensymb}
\usepackage{textcomp}
\usepackage{setspace}
\usepackage[square,comma,sort&compress,numbers]{natbib}
\usepackage{fancyhdr}
\usepackage{setspace}
\usepackage{booktabs}
\begin{document}

\onehalfspacing

\title{Diffuse scattering in metallic tin polymorphs}

\author{Bj\"orn Wehinger$^1$, Alexe\"i Bosak$^1$, Giuseppe Piccolboni$^1$, Keith Refson$^2$, Dmitry Chernyshov$^4$, Alexandre Ivanov$^4$,  Alexander Rumiantsev$^5$, and Michael Krisch$^1$}
\address{$^1$ European Synchrotron Radiation Facility, BP 220 F-38043 Grenoble Cedex 9, France}
\address{$^2$ STFC Rutherford Appleton Laboratory, Oxfordshire OX11 0QX, United Kingdom}
\address{$^3$ Swiss-Norwegian Beamlines at European Synchrotron Radiation Facility, Grenoble, France}
\address{$^4$ Institut Laue-Langevin, Grenoble, France}
\address{$^5$ Russian Academy of Sciences, Moscow, Russia}

\ead{wehinger@esrf.fr}

\begin{abstract}
The lattice dynamics of the metallic tin $\beta$ and $\gamma$ polymorphs has been studied by a combination of diffuse scattering, inelastic x-ray scattering and density functional perturbation theory. The non-symmorphic space group of the $\beta$-tin structure results in unusual asymmetry of thermal diffuse scattering. Strong resemblance of the diffuse scattering intensity distribution in $\beta$ and $\gamma$-tin were observed, reflecting the structural relationship between the two phases and revealing the qualitative similarity of the underlying electronic potential. The strong influence of the electron subsystem on inter-ionic interactions creates anomalies in the phonon dispersion relations. All observed features are described in great detail by density functional perturbation theory for both $\beta$- and $\gamma$-tin at arbitrary momentum transfers. The combined approach delivers thus a complete picture of the lattice dynamics in harmonic description. 
\end{abstract}

\maketitle

\pagestyle{fancy}
\setlength{\headheight}{16pt}
\lhead{\textit{Diffuse scattering in metallic tin polymorphs}}
\rhead{\thepage}
\cfoot{}

\section{Introduction}
Metallic tin crystallizes in a body-centred tetragonal lattice (space group I4$_1$/amd) at ambient conditions, known as white tin ($\beta-$Sn). Despite the fact that the stability range of white tin lies between 291 and $\approx$ 450 K \cite{kubiak_lcm_1986} it can be supercooled far below the transition temperature maintaining the crystal structure. Below $\approx$ 4.2 K it becomes a type-I superconductor which can be described in the frame of BCS theory \cite{matthias_rmp_1963}, indicating strong electron-phonon coupling \cite{rowe_pr_1967}. The $\alpha - \beta$ phase transition in tin is possibly the simplest and prototypical case of an entropy-driven structural transformation determined by the vibrational properties of the two phases \cite{pavone_prb_1998}.

Alloying tin with indium results in a substitutionally disordered crystal with a primitive hexagonal lattice containing one atom per unit cell \cite{raynor_am_1954}, called $\gamma$-tin. It is a convenient model system in the study of lattice dynamics and electron-phonon interactions \cite{ivanov_jpf_1987}, because its phonon dispersion relations consist only of acoustic branches and it is stable at ambient conditions. The $\gamma$-phase of pure tin \cite{kubiak_lcm_1986} is orthorhombic and differs thus slightly from the primitive hexagonal lattice. It is stable between $\approx$ 450 K and the melting point of tin (505 K).

Diffuse scattering of white tin has a long history. First Laue photographs showing a "diffuse background with regions of maximum intensities" were published in 1943 \cite{arlman_ph_1943} and "considered in the light of thermal theory" in 1946 \cite{bouman_ph_1946}. Elastic constants were derived from the diffuse features in 1955 \cite{prasad_ac_1955}. 
Phonon dispersion relations have been largely studied in the past, in particular by inelastic neutron scattering (INS) \cite{rowe_pr_1967, price_prsla_1967, rowe_prl_1965} and density functional perturbation theory \cite{pavone_prb_1998}. The available data are nevertheless limited to high-symmetry directions and the determination of eigenfrequencies. The rich Fermi surface of $\beta-$Sn \cite{devillers_pssb_1974} suggests a complex topology of electron-phonon interaction, studied in \cite{ivanov_pb_1995}.

In this study we investigate the lattice dynamics of the metallic tin polymorphs employing a combination of thermal diffuse x-ray scattering (TDS), inelastic x-ray scattering (IXS) and density functional perturbation theory (DFPT) in order to obtain the full description of the lattice dynamics at arbitrary momentum transfer.

\section{Experimental Details}
\label{sec:experiment}
The diffuse x-ray scattering experiment was conducted at the Swiss-Norwegian Beamlines at ESRF (BM01) and  the ID29 ESRF beamline. Monochromatic X-rays with wavelength 0.7 \AA \hspace{1pt} were scattered from a needle-like single crystal of $100 \times 100 $ $\mu$m cross section at room temperature. The sample was rotated orthogonal to the incoming beam over an angular range of 360$^{\circ}$ and diffuse scattering pattern were recorded in transmission geometry. Preliminary experiments were performed with a mar345 image plate detector. The follow-up experiments employed a single-photon-counting PILATUS 6M pixel detector \cite{kraft_jsr_2009}. The diffuse scattering patterns were collected in shutterless mode with a fine angular slicing of 0.1$^{\circ}$. The experimental set-up is documented elsewhere \cite{deSanctis_jsr_2012}. The orientation matrix and the geometry of the experiment were refined using the CrysAlis software package; 2D and 3D reconstructions were prepared using locally developed software.
The single crystal IXS study was carried out at beamline ID28 at the ESRF. The spectrometer was operated at an incident energy of 17.794 keV, providing an energy resolution of 3.0 meV full-width-half-maximum. IXS scans were performed in transmission geometry along selected directions in reciprocal space. Further details of the experimental set-up can be found elsewhere \cite{kirsch_Springer_2007}. 

\section{Calculation}
\label{sec:calculation}
First-principles lattice dynamics calculations were performed with the CASTEP package \cite{clark_zkri_2005, refson_prb_2006} using the DFPT solver for metallic systems \cite{degironcolo_prb_1995} at 0K. The local density approximation (LDA) and general gradient approximation (GGA) within the density functional theory formalism were used as implemented with a plane wave basis set and norm-conserving pseudopotentials. For the exchange correlation functional the Perdew and Zunger parametrization \cite{perdew_prb_1981} of the numerical results of Ceperley and Alder \cite{ceperley_LDA_prl_1980} were used in LDA and the density-gradient expansion for exchange in solids and surfaces (PBEsol functional) \cite{perdew_PBEsol_prl_2008} in GGA. 
The self-consistent electronic minimization was performed with density mixing in the Pulay scheme and the occupancies were smeared out by a Gaussian function of 0.1 eV full-width-half-maximum. The Sn pseudo-potential was of the optimized norm conserving type generated using the Vanderbilt scheme with a single projector for each of the 5s and 5p electrons, with a cut-off radius of $r_c=1.9$ $a_0$. The pseudopotentials for the LDA and PBEsol calculations were created using the CASTEP on-the-fly technology, created separately for each exchange and correlation functional and carefully tested for transferability \footnote{The CASTEP on-the-fly stings used in this work are $2|1.9|1.9|1.5|9.6|10.8|11.7|50$N$:51$N$($qc $=4.1)$ in LDA and $2|1.9|1.9|1.5|9.6|10.8|11.7|50$N$:51$N$($qc $=5.05)$ in PBEsol.}. Numerical approximations were chosen to achieve convergence to a tolerance of $ < 10^{-3}$ eV/\AA \hspace{1pt} for internal forces which required a plane wave cut-off of 380 eV and $24 \times 24 \times 24$ Monkhorst-Pack grid sampling of the first Brillouin zone.

\begin{table}[tb]
\centering
\caption{Lattice constants of $\beta$-Sn.} 
\label{tab:beta_cell}
\begin{tabular}{l l l l }
\toprule

  & LDA & PBEsol & Experiment  \cite{swanson_nist_1953} \\
  \midrule
  a = b &  5.755 \AA \hspace{1pt}  & 5.808 \AA \hspace{1pt} & 5.831  \AA \\
  c & 3.114 \AA \hspace{1pt}   & 3.144 \AA \hspace{1pt} & 3.182  \AA \\

\bottomrule

\end{tabular}
\end{table}

The structure optimization was performed using the Broyden-Fletcher-Goldfarb-Shannon method \cite{pfrommer_jcp_1997} by varying lattice and internal parameters. The equilibrium lattice constants of $\beta$-Sn as obtained in LDA and PBEsol are reported in Table \ref{tab:beta_cell}. The cell parameters agree within 2.2 \% with the experimental values in LDA and 1.2 \% in PBEsol. 
Phonon frequencies and eigenvectors were computed by perturbation calculations in harmonic approximation on a $8 \times 8 \times 8$ Monkhorst-Pack grid and further Fourier interpolated in the cumulant scheme including all image force constants \cite{parlinski_prl_1997}. The well converged internal forces yield a maximum error in phonon energies of $<$ 0.2 meV. The acoustic sum rule correction was applied to the calculated dynamical matrix in order to account for translational invariance with maximal correction of 2 meV at $\Gamma$. 

Phonon densities of states were computed using an adaptive broadening scheme for reciprocal space integration based on a method developed for electronic density of states \cite{yates_prb_2007}. In order to describe differing shapes of the contributions from steep and flat phonon branches, the states from individual branches are broadened with a Gaussian function with a broadening width 
\begin{equation}
W_{j,\bi{q}} = a \vert \frac{\partial \omega_{j,\bi{q}}}{\partial \bi{q}} \vert \Delta q
\end{equation}
proportional to the band derivatives, where $a$ is a dimensionless constant of the order of unity.

The lattice dynamics calculation for $\gamma$-tin was performed in LDA with the same parameters and pseudopotential as used for $\beta$-Sn. The primitive hexagonal structure was imposed for the unit cell containing one Sn atom. The optimized cell parameters were a = b = 3.1667 \AA \hspace{1pt} and c = 2.9722 \AA \hspace{1pt}, in agreement within 1.4 \% with the experimental values (a = b = 3.213 \AA \hspace{1pt} and c =  2.999 \AA \hspace{1pt} \cite{kubiak_bap_1974}). 
TDS and IXS intensities were calculated following the previously established formalism \cite{xu_zkri_2005,bosak_zkri_2012,kirsch_Springer_2007}, assuming  the validity of both harmonic and adiabatic approximation. The scattering intensities were calculated in first order approximation, 
\begin{equation}
\fl I (\bi{Q},\hbar \omega) = \frac{\hbar N I_e}{2} \sum_{j} \frac{1}{\omega_{\bi{Q},j}} coth \Big( \frac{\hbar \omega_{\bi{Q},j}}{2 k_B T} \Big) \vert \sum_{s} \frac{f_s}{\sqrt{m_s}} e^{-M_s} (\bi{Q} \bi{e}_{\bi{Q},j,s}) e^{-i \bi{Q} \bi{l}_s} \vert^2 \delta(\hbar \omega -\Delta \hbar \omega),
\end{equation}
where $\bi{Q}$ denotes the scattering vector, $\omega$ the phonon frequency, $\bi{e}$ the phonon eigenvector, $j$ the phonon branch index, $N$ the number of unit cells, $I_e$ the intensity from single electron scattering \cite{warren_AW_1966}, $k_B$ the Boltzmann constant, $T$ the temperature, $f$ the scattering factor of atom $s$ with mass $m$ and atomic basis vector $\bi{\l}$. The Debye Waller factor $M$ is calculated by summation over a fine wavevector ($\bi{k}$) grid,
\begin{equation}
M_s = \frac{1}{4m_s} \sum_{\bi{k},j} \frac{\hbar}{N \omega_{\bi{k},j}} coth \left( \frac{\hbar \omega_{\bi{k},j}}{2 k_B T} \right) \vert \bi{Q}  \bi{e}_{\bi{k},j,s} \vert^2.
\end{equation}

\section{Results and Discussion}
\label{sec:results}

\begin{figure}
\centering
\includegraphics[width=1.0\textwidth]{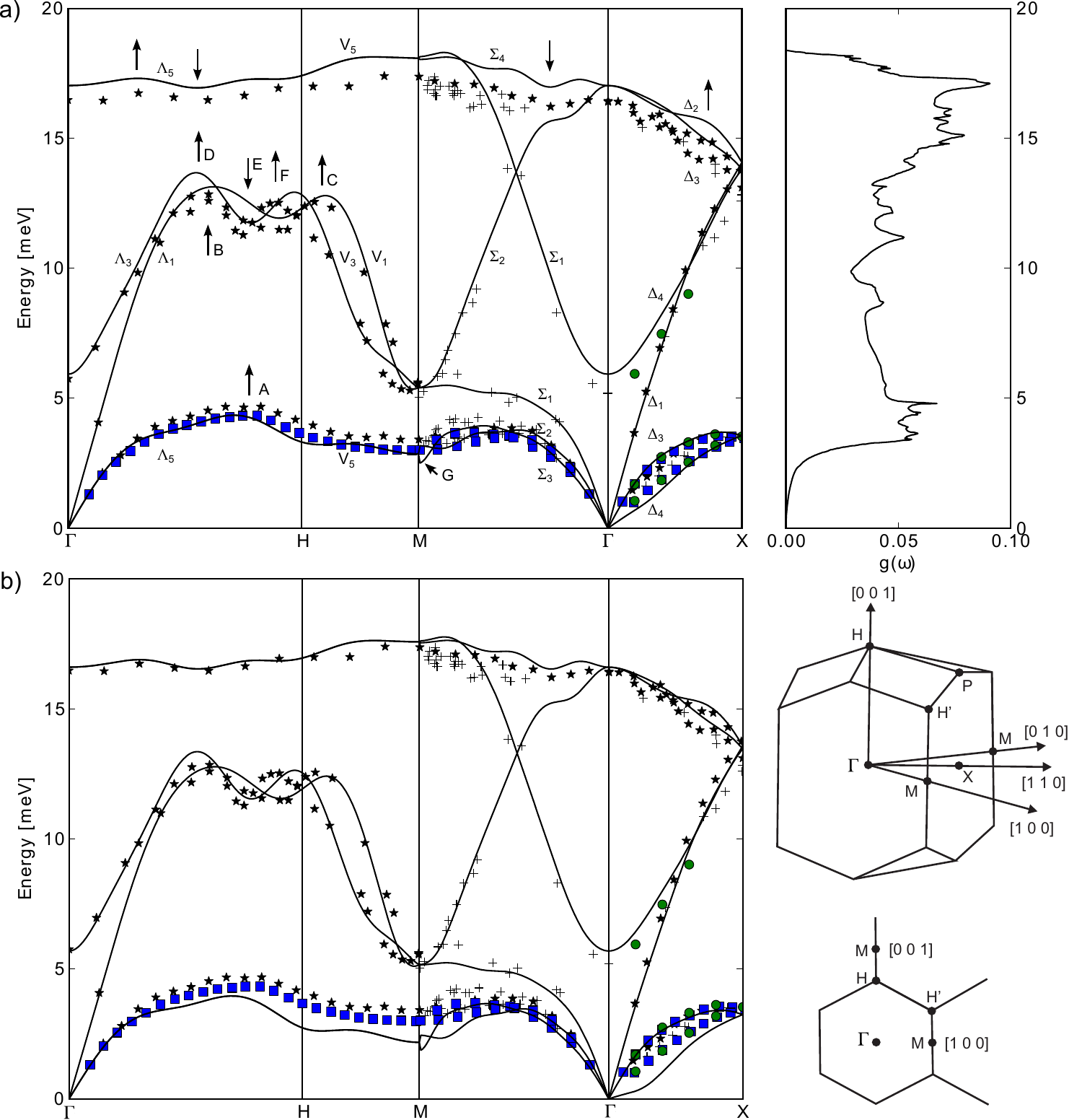}
\caption{\label{fig:b-tin_dispersion}(colour online) Dispersion relations of $\beta$-tin along the indicated high symmetry directions. The calculations (solid lines - a) LDA and b) PBEsol) are compared to experimental values from IXS measurements at 300K (circles), INS at 300K (squares) \cite{ivanov_pb_1995} and (+) \cite{price_prsla_1967} and INS at 110K ($\star$) \cite{rowe_pr_1967}. The dispersion relations along the $\Gamma$-X direction are labelled according to the symmetry classification proposed in \cite{chen_pr_1967}. The differences between the experimental data sets are due to different experimental conditions: Data were taken at different temperatures, with different statistics and resolution in momentum and energy transfer. Note the pronounced anomalies in the dispersion relations (arrows). The labelled anomalies are discussed in the text. The phonon density of states computed in LDA as well as the first Brillouin zone and a section of the H0L plane in reciprocal space are shown on the right.} 
\end{figure}

Phonon dispersion relations for $\beta$-tin along selected high symmetry directions as obtained from LDA and PBEsol calculations are presented  in Fig. \ref{fig:b-tin_dispersion} and compared to experimental results from IXS and to previously published results \cite{ivanov_jpf_1987, rowe_prl_1965,price_prsla_1967}. Both experimental and calculated dispersion relations show several anomalies, due to the complex electronic structure with long range force constants and the interplay of electrons and phonons. Some anomalies are indicated in Fig. \ref{fig:b-tin_dispersion}. The influence of the applied sum rule on the phonon branches and the anomalies was carefully tested. The transformation was found to have minimal impact on the optical branches and the anomalies, but created an artefact close to the M point, labelled G in Fig \ref{fig:b-tin_dispersion}.  Despite the fact that the lattice constants within the LDA are underestimated by the calculation, we note a good agreement for the acoustic phonon branches and the phonon anomalies. The dispersion relations are in close agreement with previous calculations by Pavone et al. \cite{pavone_prb_1998}. Some of the experimentally observed anomalies are better reproduced by the present calculation. In particular the anomalies labelled A - F are more accurately described. The highest energy optical mode shows several anomalies in both experiment and our calculation whereas the same branch is almost completely flat in the previous calculation. The anomalies are in fact sensitive to the Fermi surface which is  described more accurately in the present calculation due to a finer k-point sampling and a smaller smearing width of the occupancies.
We are therefore confident that the present calculation accounts sufficiently well for the electron-ion interaction. We note a slight over-estimation of the highest optical branch and a slight underestimation of the transverse acoustic branch in the $\Gamma$-H-M direction - the intersection of two equivalent mirror planes. A phonon with wave vector in this direction is purely longitudinal or transverse and the transverse modes are doubly degenerate. The acoustic $\Delta_4$ branch along the $\Gamma$-X direction (see Fig.  \ref{fig:b-tin_dispersion}) is slightly softer than experimentally observed. The reproducibility of the optical phonon branches is improved in the PBEsol calculation but the acoustic branch along $\Gamma$-H-M and the acoustic $\Delta_4$ branch along $\Gamma$-X are significantly softer. The low-energy branches are particular sensitive to numerical and physical errors in \textit{ab initio} lattice dynamics calculations. The case of tin is in particular difficult due to the complex electronic structure and soft character of its crystalline form. Despite the fact that the calculations are not exact, we note a good agreement in the shape of the phonon anomalies in both approximations. In the following the results of the LDA calculation were used because this study focuses on the low-energy phonons. 

\begin{figure}
\centering
\includegraphics[width=1.0\textwidth]{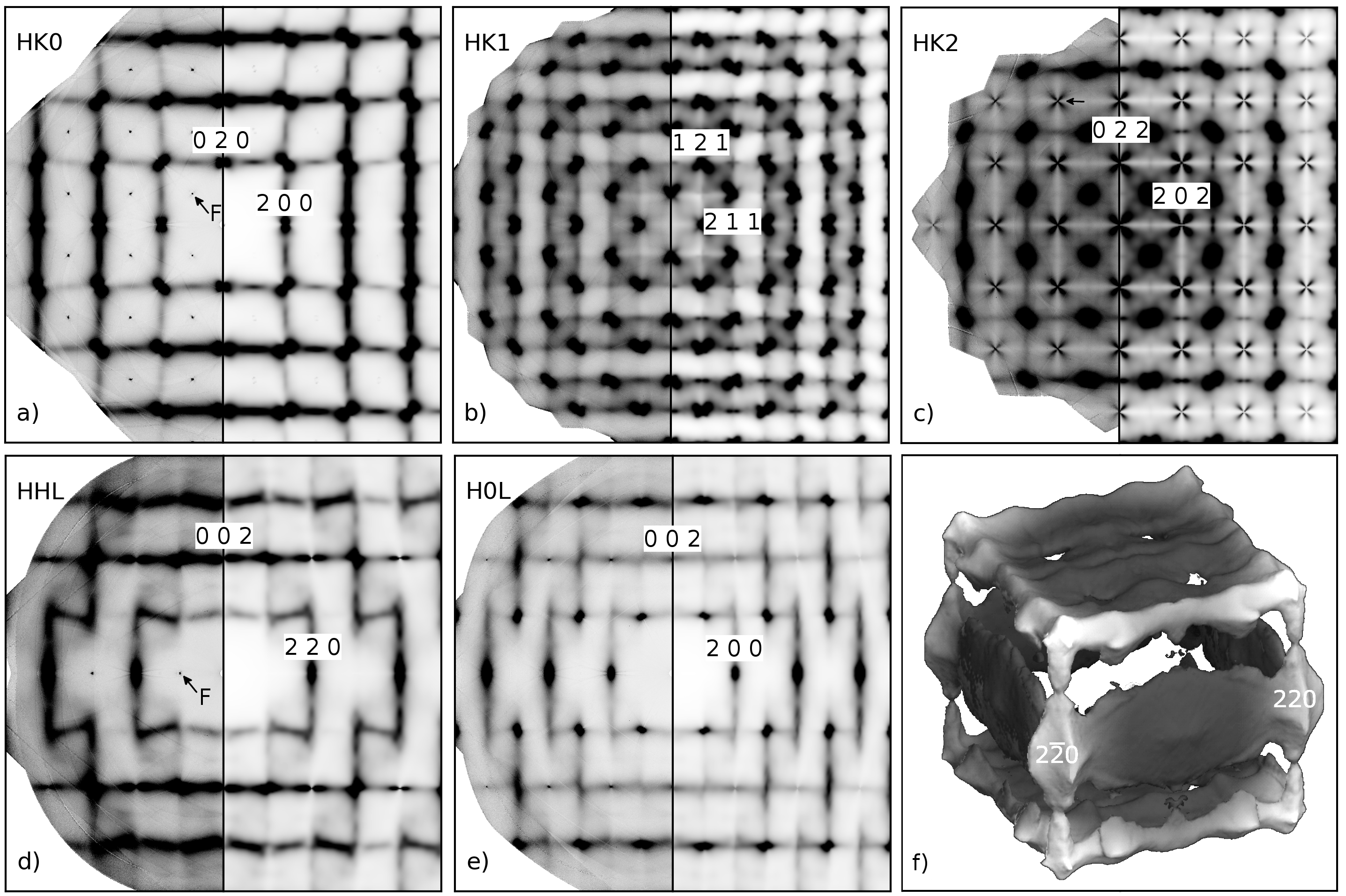}
\caption{\label{fig:TDS_planes}a) - e) Experimental diffuse scattering (left part of individual panels) and calculated (right part of individual panels) TDS intensity distribution of $\beta$-tin in the indicated reciprocal space sections. Note the almost forbidden reflections in a), c) and d) (arrows labelled F), visible due to the electron density asymmetry \cite{merisalo_prb_1979}. f) Experimental 3D isosurface of TDS  in gray scale denoting the distance from (0 0 0).} 
\end{figure}

The phonon density of states as obtained using the adaptive broadening scheme is shown beside the dispersion relation in Fig \ref{fig:b-tin_dispersion}. Adaptive broadening results in a well converged curve which is smooth at energies with highly dispersing phonon branches and captures the contribution from the phonon anomalies, which might be lost by using a conventional fixed-width broadening method. Fourier interpolating the dynamical matrices to a $48 \times 48 \times 48 $ Monkhorst-Pack grid was sufficient for convergence with a choice of $a = 1.0$. The applied first order adaptive smearing should be reliable when the energy spacing is gradient dependent. Similar to Ref. \cite{yates_prb_2007} we find that it works rather well even near critical points and the sharp features from Van Hove singularities are well described. 
We note, that the phonon density of states at very low energies is slightly affected by the underestimation of the $\Delta_4$ acoustic branch. At energies between 8 and 9 meV we find a feature in the phonon density of states which has no correlation with anomalies in the dispersion relations along high symmetry directions and must therefore originate from critical points located elsewhere.

The classical methods in the study of lattice dynamics, such as INS and IXS are flux-limited, consequently the measurements are time consuming. We therefore use TDS which allows a rapid and detailed exploration of extended regions in reciprocal space and the identification of characteristic features in the lattice dynamics. Reciprocal space sections and a three-dimensional isosurface of diffuse scattering as obtained from experiment and calculated TDS intensity distributions are shown in Fig. \ref{fig:TDS_planes}. Corrections for polarisation and projection \cite{holt_prl_1999}, and the Laue symmetry of the system were applied. All shapes of diffuse features are remarkably well reproduced by the calculation in harmonic approximation. This implies that higher order scattering processes and anharmonic effects at room temperature are much less pronounced than previously thought \cite{takahashi_ngkkk_2006}. The pronounced elastic anisotropy is reflected by the butterfly shape of TDS in the vicinity of the Brillouin zone centers $\Gamma$ in the HK0 plane. The very different sound velocities result in a large contrast in TDS intensities close to $\Gamma$ for different directions. In fact, the TDS intensities scale $ \approx 1/\omega^2$ close to $\Gamma$. 
Non-typical diffuse features are observed in the HK2n+1 and HK2n reciprocal space sections:

\begin{figure}
\centering
\includegraphics[width=0.8\textwidth]{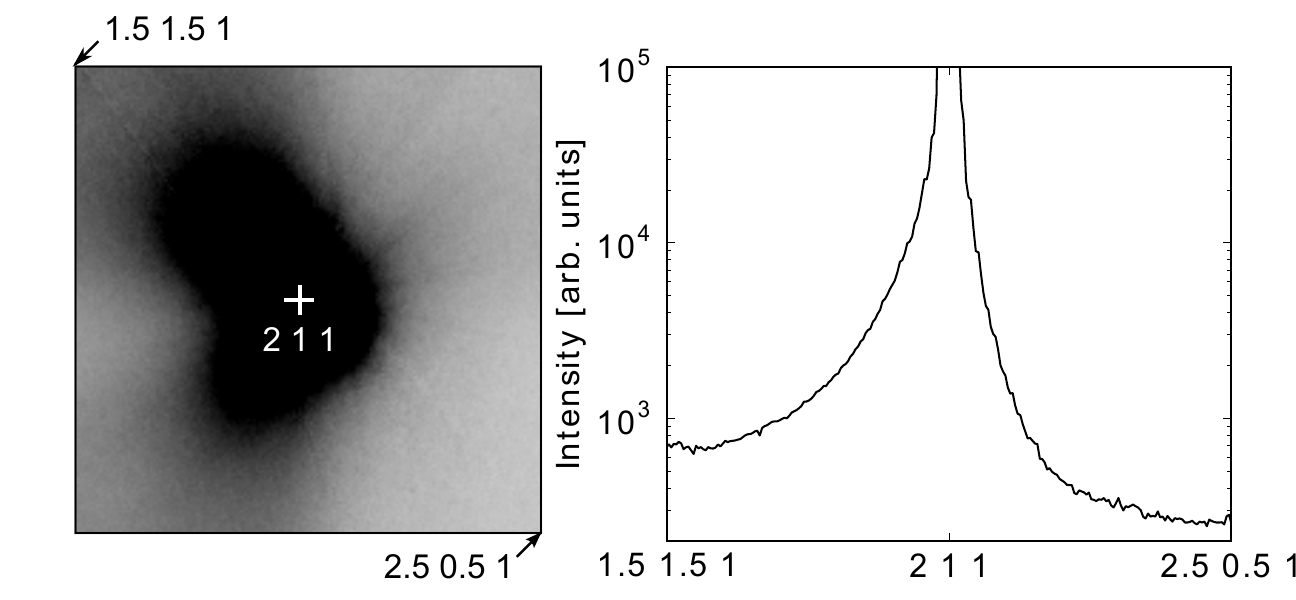}
\caption{\label{fig:211_profile}Experimental diffuse scattering of $\beta$-tin crystal in the vicinity of the 211 reflection in the HK1 plane.} 
\end{figure}

\begin{figure}
\centering
\includegraphics[width=0.8\textwidth]{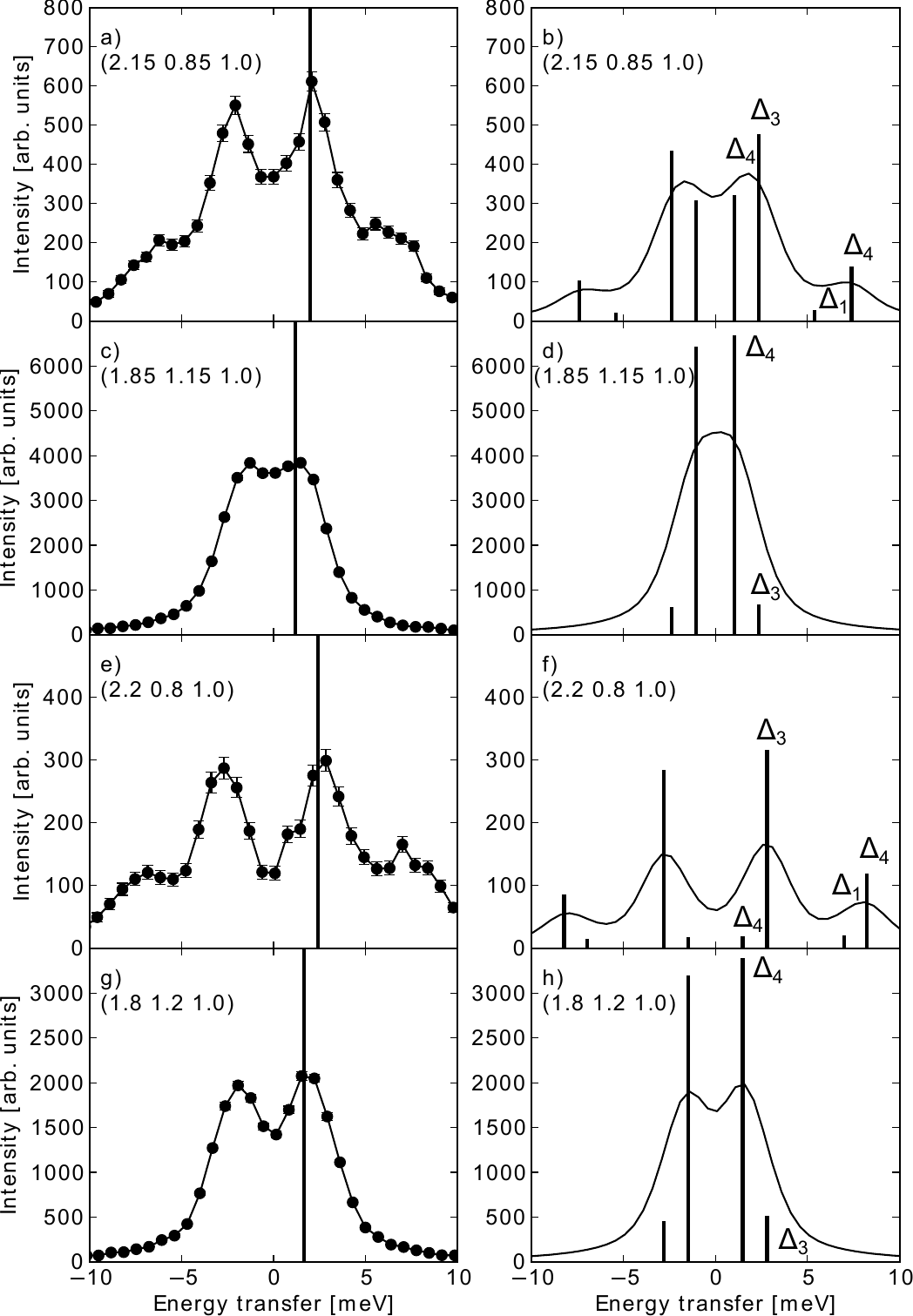}
\caption{\label{fig:ixs}Experimental IXS spectra (left panels) of $\beta$-tin on different momentum transfers in the HK1 plane. The reduced momentum transfer $q$ of panels a) and c) is equivalent, the same holds for panels e) and g). The peak position of the envelope function of the two acoustic branches is indicated by vertical lines in the experimental spectra. The inelastic contribution of the different branches (vertical lines) as obtained from the DFPT calculation and its convolution with the experimental resolution are shown for the corresponding momentum transfers in the right panels. The vertical lines are scaled by a factor 1/2 in respect to the convoluted  spectra for best visualisation.} 
\end{figure}

(i) An asymmetry in diffuse scattering of individual features is observed in the HK2n+1 reciprocal space sections. It is most pronounced around the (211) reflection, see Fig. \ref{fig:211_profile}. The HK2n+1 pattern as a whole are symmetric in agreement with the Laue symmetry of the system. The asymmetry of the diffuse scattering in the vicinity of the (211) reflections is further investigated by IXS measured at selected reduced momentum transfer $q$-values along the asymmetric TDS profile. IXS spectra are reported in Fig. \ref{fig:ixs}. Comparing the two pairs of experimental spectra at wave vectors with equivalent $q$ one observes a difference in the integrated intensities corresponding to TDS and an energy shift of the main excitation. The calculation shows that the experimentally observed excitation contains the contribution of both $\Delta_4$ and $\Delta_3$ acoustic branches. The drastic change of spectral weight between the two branches leads to an energy shift of the envelope function. The $\Delta_4$ optic branch, which is almost completely suppressed on one side, shows the same particularity.

\begin{figure}
\centering
\includegraphics[width=0.6\textwidth]{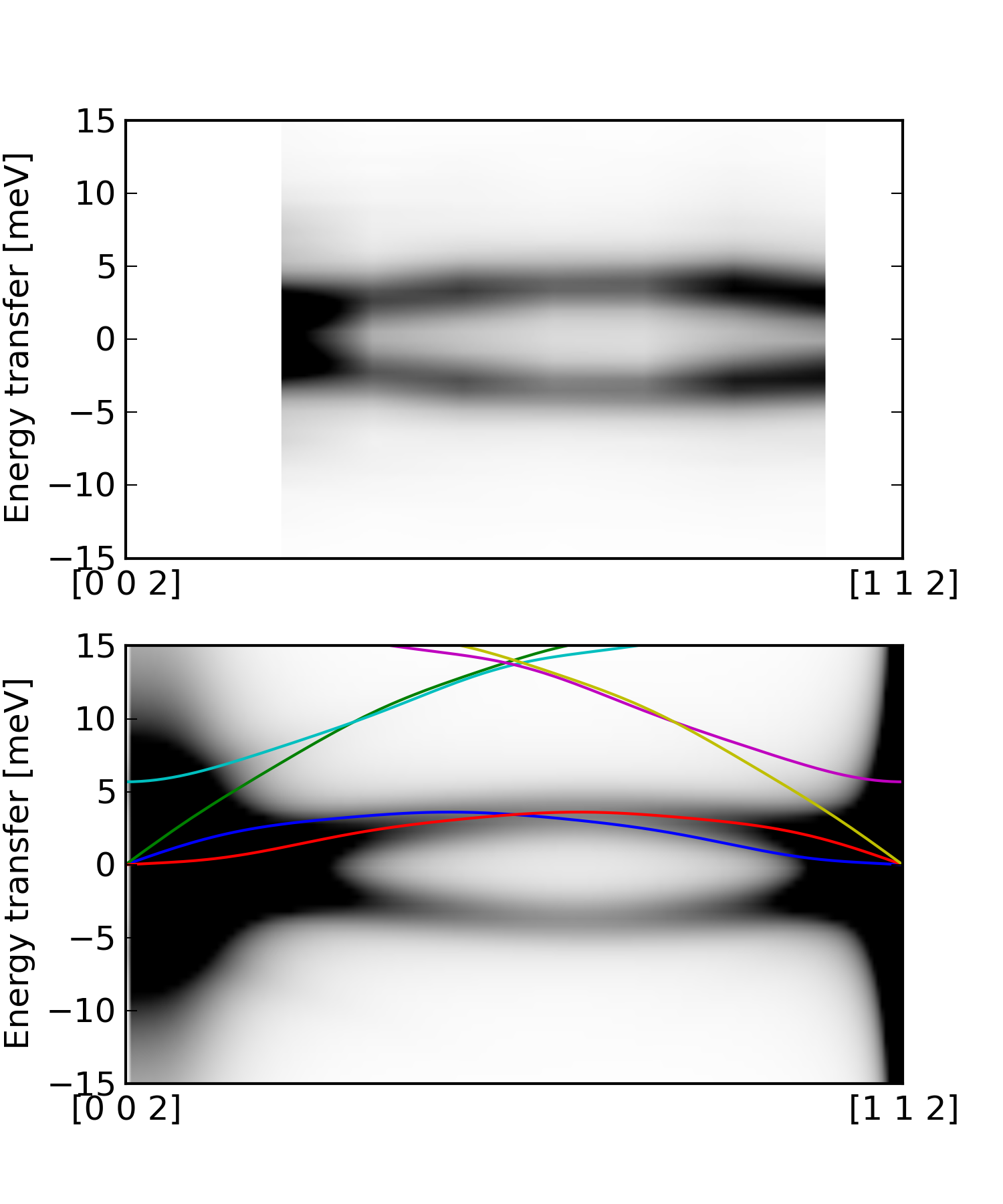}
\caption{\label{fig:ixs_maps}(colour online) IXS intensity maps from (002) to (112) as obtained from experiment (upper panel) and calculation (lower panel). The experimental map consists of eight IXS spectra with a $q$-spacing of 0.1 r.l.u and 0.7 meV energy step, linearly interpolated to a $200 \times 85$ grid. The calculated IXS intensity is convoluted with the experimental resolution function of 3.0 meV full-width-half-maximum. The dispersion of the different branches is plotted as lines.} 
\end{figure}

(ii) A cross-like feature is observed around the almost forbidden reflections in the HK2 planes (these reflections become visible due to the  asymmetry in the electron density distribution \cite{merisalo_prb_1979}). IXS is used to clarify the nature of the cross-like TDS feature. IXS scans along [$\xi$ $\xi$ 2] are summarized in an  intensity map in Fig. \ref{fig:ixs_maps}. The inelastic intensity close to (0 0 2) is dominated by the acoustic $\Delta_4$ branch with a significant contribution of the optic $\Delta_4$ branch, determined from the experiment at (0.1 0.1 2) to be 10.8 \%. The intensity close to (1 1 2) is dominated by the acoustic $\Delta_3$ branch with vanishing contribution of the optic $\Delta_4$ branch. The intensity along [$\xi$ 0 2] and [0 $\xi$ 2] is suppressed. The diffuse features in the HK2n reciprocal space sections remain symmetric.

The measured phonon energies and IXS intensities are in good agreement with the calculation in the illustrated direction. The three-dimensional isosurface of diffuse scattering intensities, depicted in Fig. \ref{fig:TDS_planes} f), allows one to identify the shape of diffuse features. We note plate-like, elongated cross-like and asymmetric shapes. The topology of the diffuse scattering is in fact quite complex and its investigation requires a fine sampling of 3D reciprocal space. The inspection of only a few planes in reciprocal space may provide an incomplete picture, the authors of a previous study \cite{takahashi_ngkkk_2006} could only identify rod-like features. 

\begin{figure}
\centering
\includegraphics[width=0.8\textwidth]{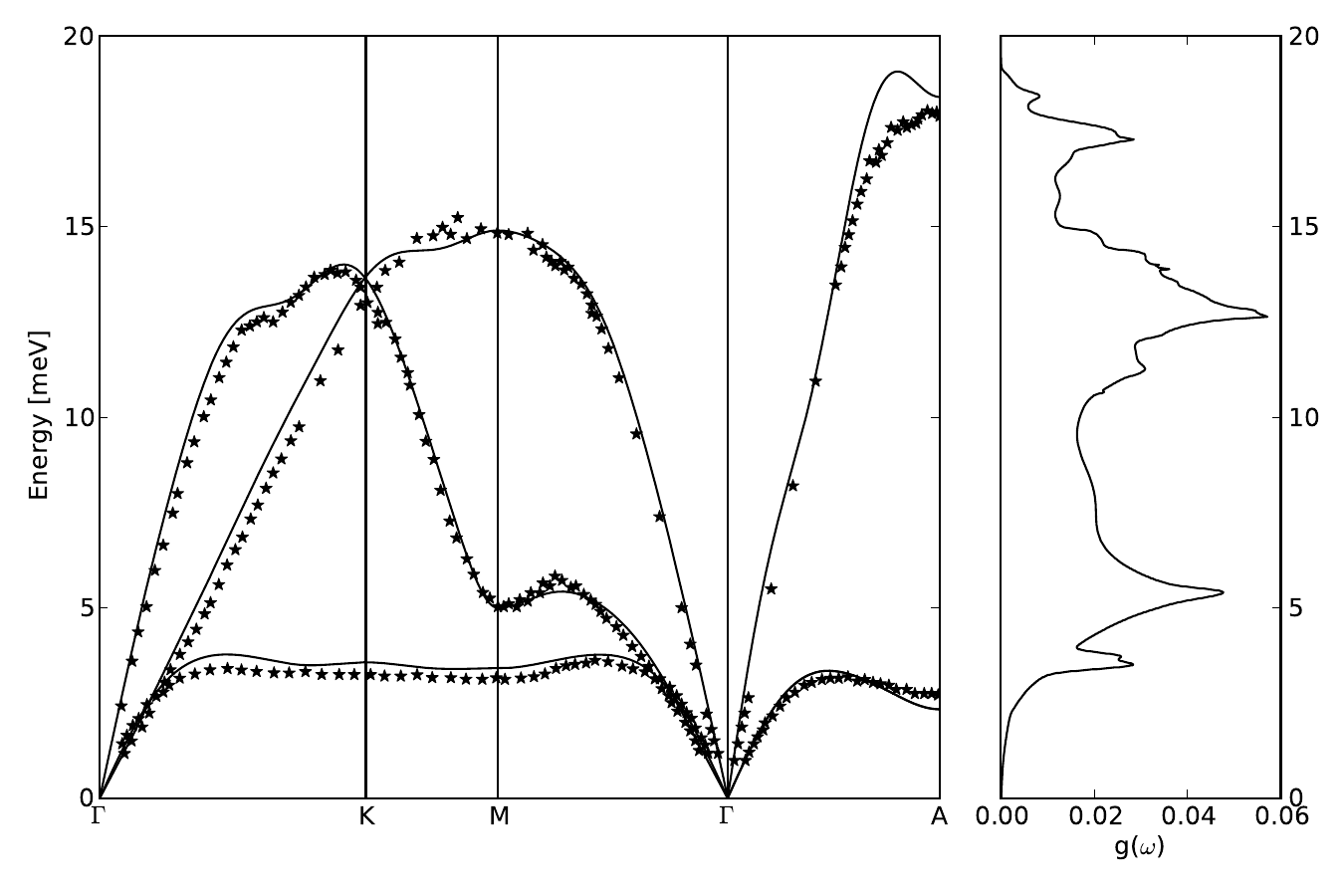}
\caption{\label{fig:g-tin_dispersion}Phonon dispersion relations along the indicated high symmetry directions and the phonon density of states of $\gamma$-tin. The calculation (solid lines, pure Sn) is compared to experimental values of a Sn$_{0.8}$In$_{0.2}$ single crystal from \cite{ivanov_jpf_1987} ($\star$). The labelled anomalies in the dispersion relation are discussed in the text.} 
\end{figure}

\begin{figure}
\centering
\includegraphics[width=0.8\textwidth]{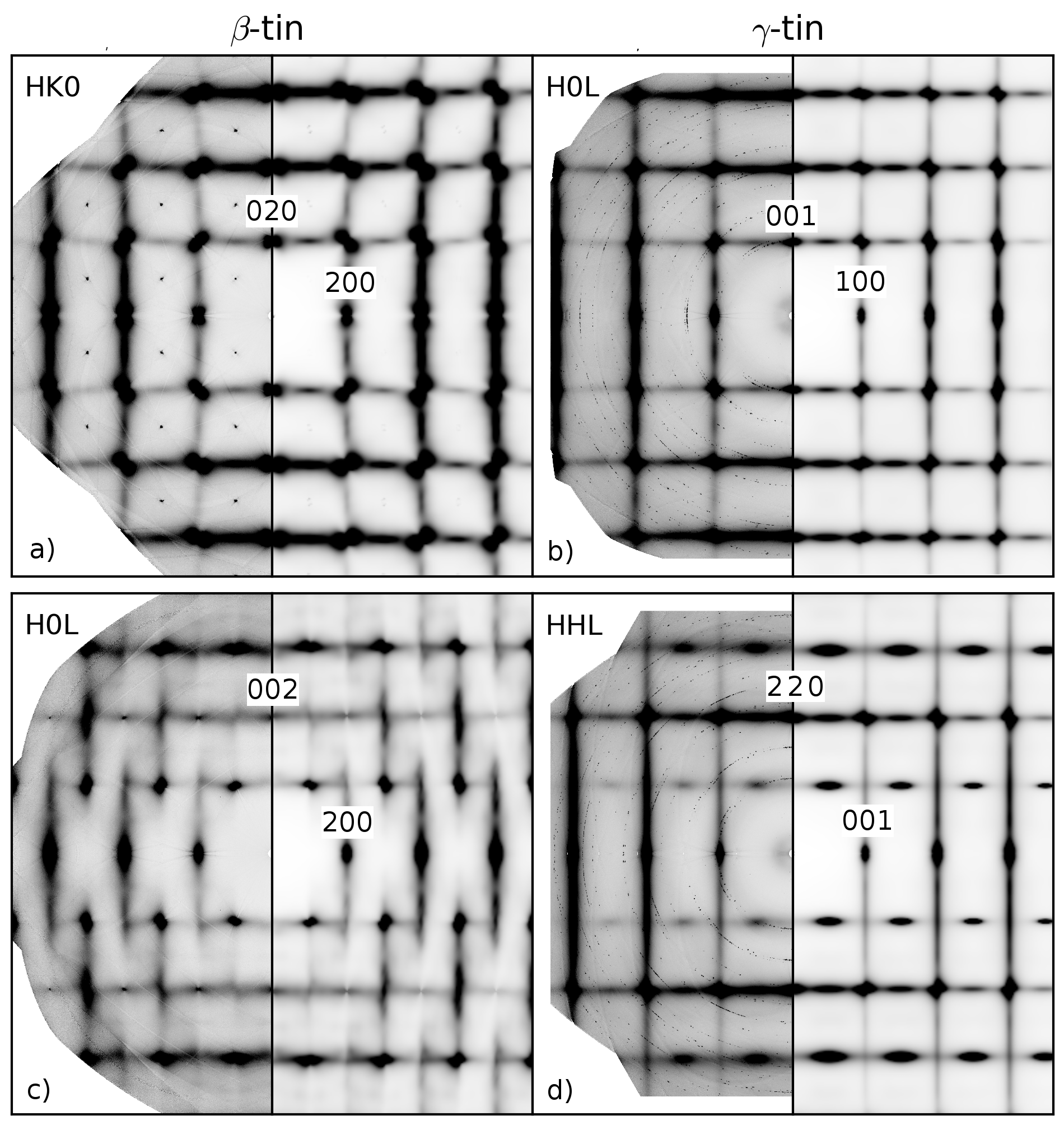}
\caption{\label{fig:TDS_beta_gamma}Experimental diffuse scattering (left part of individual panels) and calculated (right part of individual panels) TDS intensity distribution of $\beta$-tin (a) and (c) and $\gamma$-tin (b) and (d) in the indicated reciprocal space sections. The experimental TDS intensity distribution of $\gamma$-tin was obtained from a Sn$_{0.8}$In$_{0.2}$ single crystal, whereas the calculated one results from pure Sn.} 
\end{figure}

The study on $\beta$-tin was extended to $\gamma$-tin. Also the $\gamma$ phase exhibits a well defined equilibrium structure with a total static energy which is 1.85 meV per atom higher than in $\beta$-tin. Taking into account the zero-point contributions to the internal energy, the $\beta$-phase results to be more stable than the $\gamma$-phase by 0.78 meV at 0 K. The obtained dispersion relations along selected high symmetry directions and phonon density of states are presented in Fig. \ref{fig:g-tin_dispersion} and compared to experimental results from INS \cite{ivanov_jpf_1987}. We note, that the calculation describes the experimental results very well, including the phonon anomalies. Reciprocal space sections of diffuse scattering as obtained from experiment and calculated TDS are confronted to the results of $\beta$-tin in Fig. \ref{fig:TDS_beta_gamma}. The diffuse scattering in $\gamma$-tin is  almost perfectly reproduced by the calculation. The similarity of the compared TDS intensity distributions of $\beta$ and $\gamma$-tin can be appreciated. It reflects the symmetry relation of the two phases which have common subgroups, consequently some symmetry elements are retained at the $\gamma$-$\beta$ transition \cite{raynor_am_1954}. The vectorial relationship between the two structures is given by  
\begin{equation}
\left( \begin{array}{c}
\bi{a}_{\beta} \\
\bi{b}_{\beta} \\
\bi{c}_{\beta} \\
\end{array} \right) \approx \left( \begin{array}{r r r}
-1 & 1 & 0 \\
 0 & 0 & 2 \\
 1 & 1 & 0  \\
\end{array} \right)
\left( \begin{array}{c}
\bi{a}_{\gamma} \\
\bi{b}_{\gamma} \\
\bi{c}_{\gamma} \\
\end{array} \right),
\end{equation}
where the vectors $\bi{a}$, $\bi{b}$ and $\bi{c}$ denote the unit cell vectors with corresponding indices for the $\beta$ and $\gamma$ phase. 
The $\Gamma$-H-M direction in $\beta$-tin for instance corresponds to the $\Gamma$-K-M direction in $\gamma$-tin. We note, that the anomalies in the dispersion relation labelled B and C (see Fig. \ref{fig:g-tin_dispersion}) appear at the same position as in $\beta$-tin, see Fig. \ref{fig:b-tin_dispersion}. The change in slope labelled  A is less pronounced in $\gamma$-tin, resulting in a smoother intensity distribution of diffuse scattering. The structure of the two phases is different, the momentum transfer dependency of the underlying electronic potential is, however, similar. In fact the interatomic distances and force constants are very comparable in the two structures. A strong asymmetry in diffuse scattering as observed in the HK2n+1 reciprocal space sections in $\beta$-tin is not present in the $\gamma $-phase. The particularity in diffuse scattering in $\beta$-tin is thus a symmetry related feature.

\section{Conclusions}
\label{sec:conclusions}
This work demonstrates that the combination of TDS, IXS and \textit{ab initio} lattice dynamics calculation provides an optimized strategy in the study of lattice dynamics. Characteristic and anomalous features can be identified by the inspection of TDS intensity distribution in reciprocal space. Selected features were further investigated by momentum resolved IXS and confronted to the calculation. We show with experimental evidence that first principles calculations give a precise and detailed quantitative description of the lattice dynamics of the metallic tin $\beta$ and $\gamma$ polymorphs. The experimentally observed anomalies in the phonon dispersions are well reproduced by our calculation due to a very accurate description of the Fermi surface. The shapes of diffuse features are remarkably well reproduced by our calculations in harmonic approximation. This implies that higher order scattering processes and anharmonic effects at room temperature are much less pronounced than previously postulated.  An unusual asymmetry of thermal diffuse scattering is observed in $\beta$-tin which is related to the non-symmorphic structure. The comparison of TDS from $\beta$- and $\gamma$-tin reveals a strong resemblance and reflects the symmetry relation between the two structures and a strong similarity of the underlying potential.

\section*{Acknowledgments}
The authors would like to thank Daniele de Sanctis for scientific support in the diffuse scattering experiment and Alessandro Mirone and Ga\"el Goret for help in developing the software for the data analysis.


\def\newblock{\hskip .11em plus .33em minus .07em}
\bibliographystyle{unsrturl}
\bibliography{references}

\begin{thebibliography}{10}

\bibitem{kubiak_lcm_1986}
R.~Kubiak.
\newblock Evidence for the existence of the $\gamma$ form of tin.
\newblock {\em J. Less-Common Met.}, 116(2):307 -- 311, 1986.
\newblock \href {http://dx.doi.org/10.1016/0022-5088(86)90663-6}
  {\path{doi:10.1016/0022-5088(86)90663-6}}.

\bibitem{matthias_rmp_1963}
B.~T. Matthias, T.~H. Geballe, and V.~B. Compton.
\newblock Superconductivity.
\newblock {\em Rev. Mod. Phys.}, 35:1--22, Jan 1963.
\newblock \href {http://dx.doi.org/10.1103/RevModPhys.35.1}
  {\path{doi:10.1103/RevModPhys.35.1}}.

\bibitem{rowe_pr_1967}
J.~M. Rowe.
\newblock Crystal dynamics of metallic $\beta$-{S}n at 110$^\circ${K}.
\newblock {\em Phys. Rev.}, 163:547--551, Nov 1967.
\newblock \href {http://dx.doi.org/10.1103/PhysRev.163.547}
  {\path{doi:10.1103/PhysRev.163.547}}.

\bibitem{pavone_prb_1998}
P.~Pavone, S.~Baroni, and S.~de~Gironcoli.
\newblock $\alpha${}$\leftrightarrow${}$\beta${} phase transition in tin: A
  theoretical study based on density-functional perturbation theory.
\newblock {\em Phys. Rev. B}, 57:10421--10423, May 1998.
\newblock \href {http://dx.doi.org/10.1103/PhysRevB.57.10421}
  {\path{doi:10.1103/PhysRevB.57.10421}}.

\bibitem{raynor_am_1954}
G.V. Raynor and J.A. Lee.
\newblock The tin-rich intermediate phases in the alloys of tin with cadmium,
  indium and mercury.
\newblock {\em Acta Metall.}, 2(4):616 -- 620, 1954.
\newblock \href {http://dx.doi.org/10.1016/0001-6160(54)90197-2}
  {\path{doi:10.1016/0001-6160(54)90197-2}}.

\bibitem{ivanov_jpf_1987}
A.S. Ivanov, A.Y. Rumiantsev, B.~Dorner, N.L. Mitrofanov, and V.V. Pushkarev.
\newblock Lattice dynamics and electron-phonon interaction in $\gamma$-tin.
\newblock {\em J. Phys. F: Met. Phys.}, 17(9):1925, 1987.
\newblock \href {http://dx.doi.org/10.1088/0305-4608/17/9/017}
  {\path{doi:10.1088/0305-4608/17/9/017}}.

\bibitem{arlman_ph_1943}
J.J. Arlman and R.~Kronig.
\newblock Investigation of lattice defects by means of {X}-rays. {I}. {T}in.
\newblock {\em Physica}, 10(10):795 -- 800, 1943.
\newblock \href {http://dx.doi.org/10.1016/S0031-8914(43)80003-3}
  {\path{doi:10.1016/S0031-8914(43)80003-3}}.

\bibitem{bouman_ph_1946}
J.~Bouman, J.J. Arlman, and L.L. Van~Reijen.
\newblock Investigations of lattice defects by means of {X}-rays: {IV}. the
  diffuse pattern of tin single crystals.
\newblock {\em Physica}, 12(6):353 -- 370, 1946.
\newblock \href {http://dx.doi.org/10.1016/S0031-8914(46)80055-7}
  {\path{doi:10.1016/S0031-8914(46)80055-7}}.

\bibitem{prasad_ac_1955}
S.~C. Prasad and W.~A. Wooster.
\newblock {The study of the elastic constants of white tin by diffuse x-ray
  reflexion}.
\newblock {\em Acta Crystallogr.}, 8(11):682--686, Nov 1955.
\newblock \href {http://dx.doi.org/10.1107/S0365110X55002119}
  {\path{doi:10.1107/S0365110X55002119}}.

\bibitem{price_prsla_1967}
L.G. Parratt.
\newblock Lattice dynamics of white tin.
\newblock {\em Proc. R. Soc. Lond. A}, 300:25--44, 1967.

\bibitem{rowe_prl_1965}
J.~M. Rowe, B.~N. Brockhouse, and E.~C. Svensson.
\newblock Lattice dynamics of white tin.
\newblock {\em Phys. Rev. Lett.}, 14:554--556, Apr 1965.
\newblock \href {http://dx.doi.org/10.1103/PhysRevLett.14.554}
  {\path{doi:10.1103/PhysRevLett.14.554}}.

\bibitem{devillers_pssb_1974}
M.~A.~C. Deviller, M.~M. M.~P. Matthey, and A.~R. De~Vroomen.
\newblock Fermi-surface of white tin from a {RAPW} interpolation compared with
  experiment.
\newblock {\em Phys. Status Solidi B}, 63(2):471--484, 1974.
\newblock \href {http://dx.doi.org/10.1002/pssb.2220630206}
  {\path{doi:10.1002/pssb.2220630206}}.

\bibitem{ivanov_pb_1995}
A.S. Ivanov, N.L. Mitrofanov, and A.Y. Rumiantsev.
\newblock Fermi surface and fine structure of the phonon dispersion curves of
  white tin.
\newblock {\em Phys. B}, 213–214(0):423 -- 426, 1995.
\newblock \href {http://dx.doi.org/10.1016/0921-4526(95)00177-B}
  {\path{doi:10.1016/0921-4526(95)00177-B}}.

\bibitem{kraft_jsr_2009}
P.~Kraft, A.~Bergamaschi, Ch. Broennimann, R.~Dinapoli, E.~F. Eikenberry,
  B.~Henrich, I.~Johnson, A.~Mozzanica, C.~M. Schlep{\"{u}}tz, P.~R. Willmott,
  and B.~Schmitt.
\newblock Performance of single-photon-counting {PILATUS} detector modules.
\newblock {\em J. Synchrotron Radiat.}, 16(3):368--375, May 2009.
\newblock \href {http://dx.doi.org/10.1107/S0909049509009911}
  {\path{doi:10.1107/S0909049509009911}}.

\bibitem{deSanctis_jsr_2012}
D.~de~Sanctis, A.~Beteva, H.~Caserotto, F.~Dobias, J.~Gabadinho, T.~Giraud,
  A.~Gobbo, M.~Guijarro, M.~Lentini, B.~Lavault, T.~Mairs, S.~McSweeney,
  S.~Petitdemange, V.~Rey-Bakaikoa, J.~Surr, P.~Theveneau, G.~Leonard, and
  C.~Mueller-Dieckmann.
\newblock {{ID}29: a high-intensity highly automated {ESRF} beamline for
  macromolecular crystallography experiments exploiting anomalous scattering}.
\newblock {\em J. Synchrotron Radiat.}, 19(3):455--461, May 2012.
\newblock \href {http://dx.doi.org/10.1107/S0909049512009715}
  {\path{doi:10.1107/S0909049512009715}}.

\bibitem{kirsch_Springer_2007}
M.~Krisch and F.~Sette.
\newblock {\em Inelastic X-ray Scattering from Phonons. Light Scattering in
  solids, Novel Materials and Techniques, Topics in Applied Physics 108}.
\newblock Springer-Verlag, Berlin Heidelberg, 2007.

\bibitem{clark_zkri_2005}
S.~Clark, M.~Segall, C.~Pickard, P.~Hasnip, M.~Probert, K.~Refson, and
  M.~Payne.
\newblock First principles methods using {CASTEP}.
\newblock {\em Z. Kristallogr.}, 220:567--570, 2005.
\newblock \href {http://dx.doi.org/10.1524/zkri.220.5.567.65075}
  {\path{doi:10.1524/zkri.220.5.567.65075}}.

\bibitem{refson_prb_2006}
Keith Refson, Paul~R. Tulip, and Stewart~J. Clark.
\newblock Variational density-functional perturbation theory for dielectrics
  and lattice dynamics.
\newblock {\em Phys. Rev. B}, 73:155114, Apr 2006.
\newblock \href {http://dx.doi.org/10.1103/PhysRevB.73.155114}
  {\path{doi:10.1103/PhysRevB.73.155114}}.

\bibitem{degironcolo_prb_1995}
S.~de~Gironcoli.
\newblock Lattice dynamics of metals from density-functional perturbation
  theory.
\newblock {\em Phys. Rev. B}, 51:6773--6776, Mar 1995.
\newblock \href {http://dx.doi.org/10.1103/PhysRevB.51.6773}
  {\path{doi:10.1103/PhysRevB.51.6773}}.

\bibitem{perdew_prb_1981}
J.~P. Perdew and A.~Zunger.
\newblock Self-interaction correction to density-functional approximations for
  many-electron systems.
\newblock {\em Phys. Rev. B}, 23:5048--5079, May 1981.
\newblock \href {http://dx.doi.org/10.1103/PhysRevB.23.5048}
  {\path{doi:10.1103/PhysRevB.23.5048}}.

\bibitem{ceperley_LDA_prl_1980}
D.~M. Ceperley and B.~J. Alder.
\newblock Ground state of the electron gas by a stochastic method.
\newblock {\em Phys. Rev. Lett.}, 45:566--569, Aug 1980.
\newblock \href {http://dx.doi.org/10.1103/PhysRevLett.45.566}
  {\path{doi:10.1103/PhysRevLett.45.566}}.

\bibitem{perdew_PBEsol_prl_2008}
J.~P. Perdew, A.~Ruzsinszky, G.~I. Csonka, O.~A. Vydrov, G.~E. Scuseria, L.~A.
  Constantin, X.~Zhou, and K.~Burke.
\newblock Restoring the density-gradient expansion for exchange in solids and
  surfaces.
\newblock {\em Phys. Rev. Lett.}, 100:136406, 2008.
\newblock \href {http://dx.doi.org/10.1103/PhysRevLett.100.136406}
  {\path{doi:10.1103/PhysRevLett.100.136406}}.

\bibitem{pfrommer_jcp_1997}
B.~G. Pfrommer, M.~Côté, S.~G. Louie, and M.~L. Cohen.
\newblock Relaxation of crystals with the quasi-newton method.
\newblock {\em J. Comput. Phys.}, 131(1):233 -- 240, 1997.
\newblock \href {http://dx.doi.org/10.1006/jcph.1996.5612}
  {\path{doi:10.1006/jcph.1996.5612}}.

\bibitem{parlinski_prl_1997}
K.~Parlinski, Z.~Q. Li, and Y.~Kawazoe.
\newblock First-principles determination of the soft mode in cubic {Z}r{O}$_2$.
\newblock {\em Phys. Rev. Lett.}, 78:4063--4066, May 1997.
\newblock \href {http://dx.doi.org/10.1103/PhysRevLett.78.4063}
  {\path{doi:10.1103/PhysRevLett.78.4063}}.

\bibitem{yates_prb_2007}
Jonathan~R. Yates, Xinjie Wang, David Vanderbilt, and Ivo Souza.
\newblock Spectral and fermi surface properties from wannier interpolation.
\newblock {\em Phys. Rev. B}, 75:195121, May 2007.
\newblock \href {http://dx.doi.org/10.1103/PhysRevB.75.195121}
  {\path{doi:10.1103/PhysRevB.75.195121}}.

\bibitem{kubiak_bap_1974}
R.~Kubiak and Lukaszew. K.
\newblock Crystal-structure and thermal-expansion of {I}n$_3${S}n and
  {I}n{S}n$_4$.
\newblock {\em Bull. Acad. Pol. Sci., Ser. Sci. Chim.}, 22(4):281--286, 1974.

\bibitem{xu_zkri_2005}
R.Q. Xu and T.C. Chiang.
\newblock Determination of phonon dispersion relations by x-ray thermal diffuse
  scattering.
\newblock {\em Z. Kristallogr.}, 220(12):1009--1016, 2005.
\newblock \href {http://dx.doi.org/10.1524/zkri.2005.220.12.1009}
  {\path{doi:10.1524/zkri.2005.220.12.1009}}.

\bibitem{bosak_zkri_2012}
A.~Bosak, M.~Krisch, D.~Chernyshov, B.~Winkler, V.~Milman, K.~Refson, and
  C.~Schulze-Briese.
\newblock New insights into the lattice dynamics of alpha-quartz.
\newblock {\em Z. Kristallogr.}, 227(2):84--91, 2012.
\newblock \href {http://dx.doi.org/10.1524/zkri.2012.1432}
  {\path{doi:10.1524/zkri.2012.1432}}.

\bibitem{warren_AW_1966}
B.~E. Warren.
\newblock {\em X-ray Diffraction}.
\newblock Addison-Weley, Reading, 1966.

\bibitem{holt_prl_1999}
M.~Holt, Z.~Wu, Hawoong Hong, P.~Zschack, P.~Jemian, J.~Tischler, Haydn Chen,
  and T.-C. Chiang.
\newblock Determination of phonon dispersions from x-ray transmission
  scattering: The example of silicon.
\newblock {\em Phys. Rev. Lett.}, 83:3317--3319, Oct 1999.
\newblock \href {http://dx.doi.org/10.1103/PhysRevLett.83.3317}
  {\path{doi:10.1103/PhysRevLett.83.3317}}.

\bibitem{takahashi_ngkkk_2006}
M.~Takahashi, K.~Ohshima, and Y.~Noda.
\newblock Measurement of diffuse scattering in $\beta$-tin single crystal.
\newblock {\em Nihon Genshiryoku Kenkyu Kaihatsu Kiko JAEA- Review}, 2006.

\bibitem{merisalo_prb_1979}
M.~Merisalo and J.~Soininen.
\newblock Covalency of bonding in $\beta$-{S}n.
\newblock {\em Phys. Rev. B}, 19:6289--6294, Jun 1979.
\newblock \href {http://dx.doi.org/10.1103/PhysRevB.19.6289}
  {\path{doi:10.1103/PhysRevB.19.6289}}.

\bibitem{swanson_nist_1953}
H.E. Swanson and E.~Tatge.
\newblock Standard x-ray diffraction powder patterns.
\newblock {\em Natl. Bur. Stand.}, 539:1--95, 1953.

\bibitem{chen_pr_1967}
S.~H. Chen.
\newblock Group-theoretical analysis of lattice vibrations in metallic
  $\beta$-{S}n.
\newblock {\em Phys. Rev.}, 163:532--546, Nov 1967.
\newblock \href {http://dx.doi.org/10.1103/PhysRev.163.532}
  {\path{doi:10.1103/PhysRev.163.532}}.

\end{thebibliography}

\clearpage

\end{document}